\documentclass{PoS}

\title{Analytical continuation from imaginary to real chemical potential
in 2-color QCD under scrutiny}

\ShortTitle{Analytical continuation from imaginary $\mu$ under scrutiny}

\author{\speaker{Alessandro Papa}\\
        Universit\`a della Calabria \& INFN-Cosenza\\
        E-mail: \email{papa@cs.infn.it}}

\author{Paolo Cea\\
        Universit\`a di Bari \& INFN-Bari\\
        E-mail: \email{paolo.cea@ba.infn.it}}

\author{Leonardo Cosmai\\
        INFN-Bari\\
        E-mail: \email{leonardo.cosmai@ba.infn.it}}

\author{Massimo D'Elia\\
        Universit\`a di Genova \& INFN-Genova\\
        E-mail: \email{Massimo.Delia@ge.infn.it}}

\abstract{The method of analytical continuation from imaginary to real chemical
potential is tested in 2-color QCD. In comparison to previous studies in the
same theory, an exact updating algorithm is used and simulations are 
performed closer to the thermodynamic limit. It is shown that the 
method considerably improves if suitable functions are used to interpolate
data with imaginary chemical potential.}

\FullConference{XXIV International Symposium on Lattice Field Theory\\
		 July 23-28 2006\\
		 Tucson Arizona, US}

\begin{document}

\section{Introduction}

The identification of the phases of QCD in the temperature -- chemical potential plane and the 
exact localization and nature of the transitions among them are of central interest nowadays,
due to the implications in cosmology, in astrophysics and in the phenomenology of heavy ion 
collisions. Most aspects of the QCD phase diagram are beyond the reach of perturbation theory and 
the lattice approach is a natural tool to face them. For non-zero chemical potential,
however, the QCD fermion determinant becomes complex and the standard Monte Carlo importance
sampling is unfeasible -- the well-known ``sign problem''.

Several strategies have been invented to circumvent this problem (for a review, see~\cite{Phi05}
and~\cite{Sch06}) and some useful information on the critical line separating the hadronic phase from
the quark-gluon plasma phase in the region $\mu/T\lesssim 1$ has been already achieved. 

Here, we concentrate on one of these approaches, the method of analytical continuation,
first used in Ref.~\cite{Lom00} and in Ref.~\cite{HLP01}. The idea behind this method is very simple:
numerical simulations are performed at {\it imaginary} chemical potential, $\mu=i\mu_I$,
for which the fermion 
determinant is real, then Monte Carlo determinations are interpolated by a suitable function and
finally this function is analytically continued to real values of $\mu$. This method is rather 
powerful since $\beta$ and $\mu$ parameters can be varied independently and there is no
limitation from increasing lattice size, as in methods based on reweighting. There is, however,
an important drawback: the periodicity of the partition function and the presence 
of non-analyticities arising for imaginary values of the chemical potential~\cite{RW86} make 
that the region useful 
for numerical determinations is inside the strip $0\leq\mu_I/T<\pi/3$. This implies that the 
accuracy in the interpolation of the results at imaginary chemical potential has a strong impact
on the extension of the domain of real $\mu$ values reachable after analytical continuation.

So far, the method of analytical continuation has been applied in SU(3) with $n_f=2$~\cite{FP02},
$n_f=3$~\cite{FP03} and $n_f=4$~\cite{DL}. Moreover, it has been tested in several
theories which do not suffer the sign problem, by direct comparison of the analytical continuation
with Monte Carlo results obtained at real $\mu$~\cite{HLP01,GP04,Kim05}.

In all these applications a truncated Taylor series (or, more simply, a polynomial) has been used 
as interpolating function. In this work we show that the method can be considerably improved if 
ratios of polynomials are used instead. As a test-field for our proposal we adopt the same
theory used in Ref.~\cite{GP04}, namely SU(2) (or 2-color QCD), with $n_f=8$ degenerate staggered 
fermions.

\section{Theoretical background}

\begin{figure}
\includegraphics[width=0.5\textwidth]{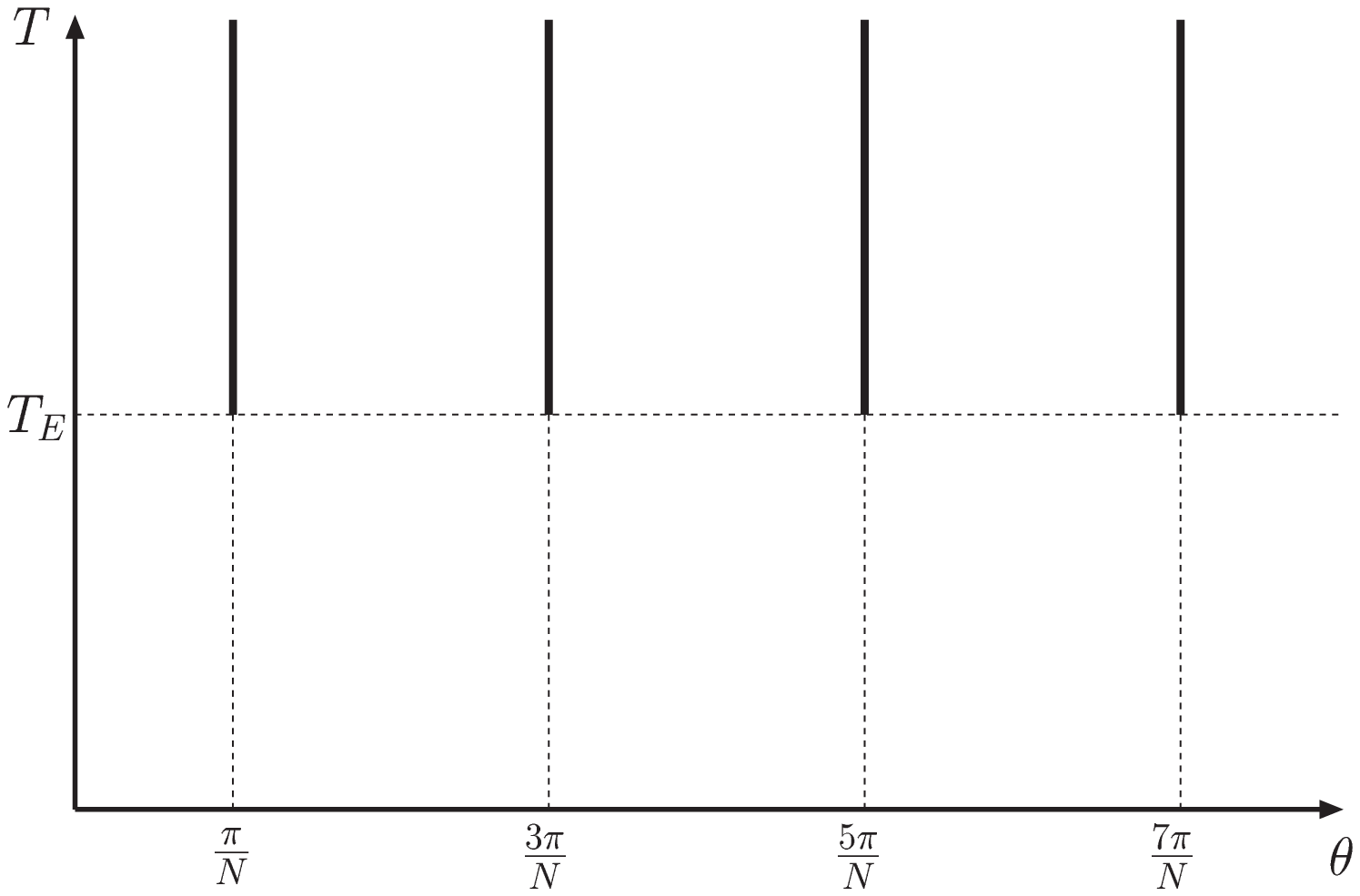}
\includegraphics[width=0.5\textwidth]{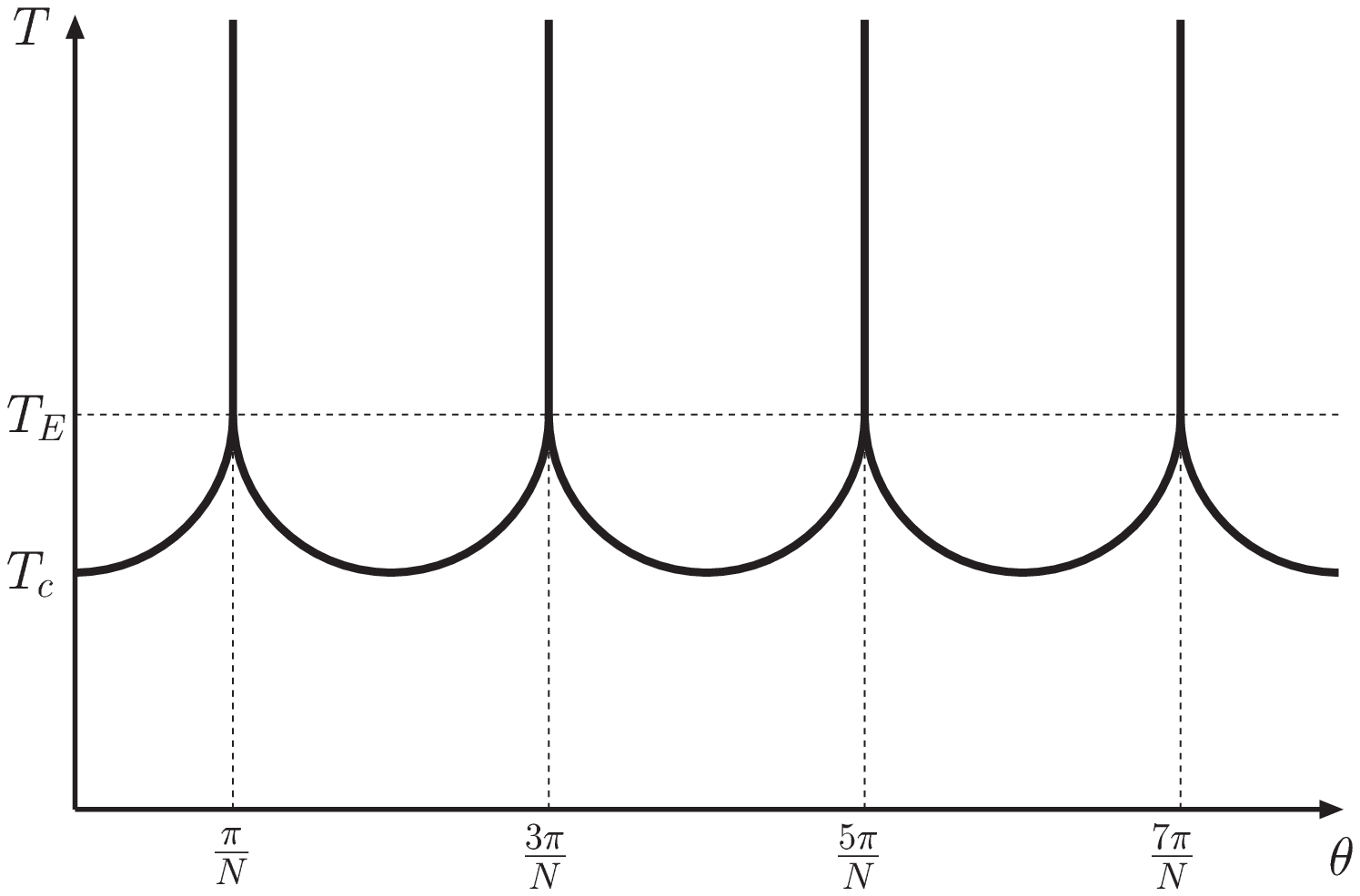}
\caption[]{(Left) Phase diagram in the $(T,\theta)$ plane according to Ref.~\cite{RW86}.
(Right) Tentative phase diagram in the $(T,\theta)$ plane after the inclusion of the chiral 
critical lines.}
\label{fig1}
\end{figure}

Long ago Roberge and Weiss have shown~\cite{RW86} that the partition function of 
any SU($N$) gauge theory with non-zero temperature and imaginary chemical potential, 
$\mu=i\mu_I$, is periodic in $\theta\equiv\mu_I/T$ with period $2\pi/N$ and 
that the free energy $F$ is a regular function of
$\theta$ for $T < T_E$, while it is discontinuous at $\theta=2\pi(k+1/2)/N$, $k=0,1,2,\ldots$,
for $T > T_E$, where $T_E$ is a characteristic temperature, depending on the theory. The resulting
phase diagram in the $(T,\theta)$-plane is given in Fig.~\ref{fig1} (left), where the
vertical lines represent first order transition lines. This structure is compatible with the 
$\mu\to -\mu$ symmetry, related with CP invariance, and with the Roberge-Weiss
periodicity. The $\mu_I$-dependence of any observable is completely determined if this
observable is known in the strip $0\leq \theta < \pi/N$.  These predictions have been confirmed 
numerically in several cases, studying the behaviour of quantities like the Polyakov loop and 
the chiral condensate~\cite{FP02,DL,GP04}. 

A phase diagram like that in Fig.~\ref{fig1} (left) would imply the absence of any transition along 
the $T$ axis in the physical regime of zero chemical potential for any value of $N$, of $n_f$
and of the quark masses, which cannot be true. Therefore, it is necessary to admit that the phase
diagram in the $(T,\theta)$-plane is more complicated than in Fig.~\ref{fig1} (left). 
The simplest possibility is given in Fig.~\ref{fig1} (right), where the added lines generally
represent transitions which can be first order, second order or crossover. The temperature
$T_c$ is the critical or pseudo-critical one for the transition at zero chemical potential.
It is convenient to redraw the phase diagram of Fig.~\ref{fig1} (right) in the 
$(\beta,\hat \mu_I)$-plane (Fig.~\ref{fig2}),
where $\beta=2 N/g^2$, $\hat \mu_I=a \mu_I$ is the imaginary chemical potential in lattice units
and it has been used the fact that $T=1/(a N_\tau)$, with $N_\tau$ the temporal extension of the 
lattice.

\section{Numerical results}

\begin{figure}
\begin{center}
\includegraphics[width=0.65\textwidth]{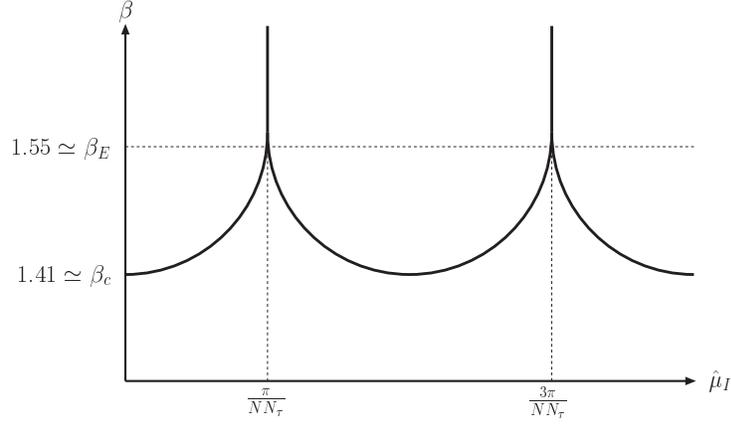}
\caption[]{Phase diagram in the $(\beta,\hat\mu_I)$-plane; $N$ is the number of colors, $N_\tau$
the extension of the lattice in the temporal direction. The numerical values for $\beta_E$
and $\beta_c$ are valid for SU(2) in presence of $n_f=8$ degenerate staggered fermions with 
mass $am=0.07$.}
\label{fig2}
\end{center}
\end{figure}

We performed numerical simulations on a $16^3\times 4$ lattice of the SU(2) gauge theory with 
$n_f=8$ degenerate staggered fermions having mass $am=0.07$. For this theory the tentative 
phase diagram looks like in Fig.~\ref{fig2}, with $\beta_E\simeq 1.55$~\cite{GP04} 
and $\beta_c\simeq1.41$\cite{LMNT00}. 
We used the hybrid Monte Carlo algorithm, with $dt=0.01$. The observables we determined are the 
Polyakov loop, the chiral condensate and the fermionic number density. We have chosen $\beta=1.90>\beta_E$
and have taken 1000-5000 measurements for several values of the imaginary chemical potential
$\mu_I$ in the interval from zero to the value corresponding to the first RW transition line, 
$\hat\mu_I=\pi/8$, and for several values of the real chemical potential $\mu_R$ in the
interval $[0,2]$. 
Simulations have been performed on the APEmille crate in Bari and on the recently 
installed computer facilities at the INFN apeNEXT Computing Center in Rome.

We have used the data at imaginary chemical potential $\mu_I$ to determine the parameters of the
interpolating function, then we have analytically continued this function to real values of the
chemical potential and compared there with direct Monte Carlo determinations. In order to fulfill 
CP invariance,
the interpolating function must be a function of $\mu^2$ for observables, such as the 
Polyakov loop and the chiral condensate, which do not depend explicitly on $\mu$. The fermion
number density, being the logarithmic derivative of the partition function with respect to the
chemical potential, is instead an odd function of $\mu$. For the Polyakov loop
and the chiral condensate we have considered a second order polynomial in $\mu^2$, 
\begin{equation}
A+B \mu^2+C \mu^4\;,
\label{poly}
\end{equation}
according to the standard approach, and the ratio of two first oder polynomials in $\mu^2$,
\begin{equation}
\frac{A+B \mu^2}{1+C \mu^2}\;,
\label{ratio}
\end{equation}
according to our new proposal.
Similarly, for the fermionic number density we have used a polynomial of the form
\begin{equation}
A \mu+B \mu^3+C \mu^5\;,
\end{equation}
and the ratio
\begin{equation}
\frac{A\mu+B \mu^3}{1+C \mu^2}\;.
\end{equation}

\begin{figure}
\begin{center}
\includegraphics[width=0.75\textwidth,angle=-90]{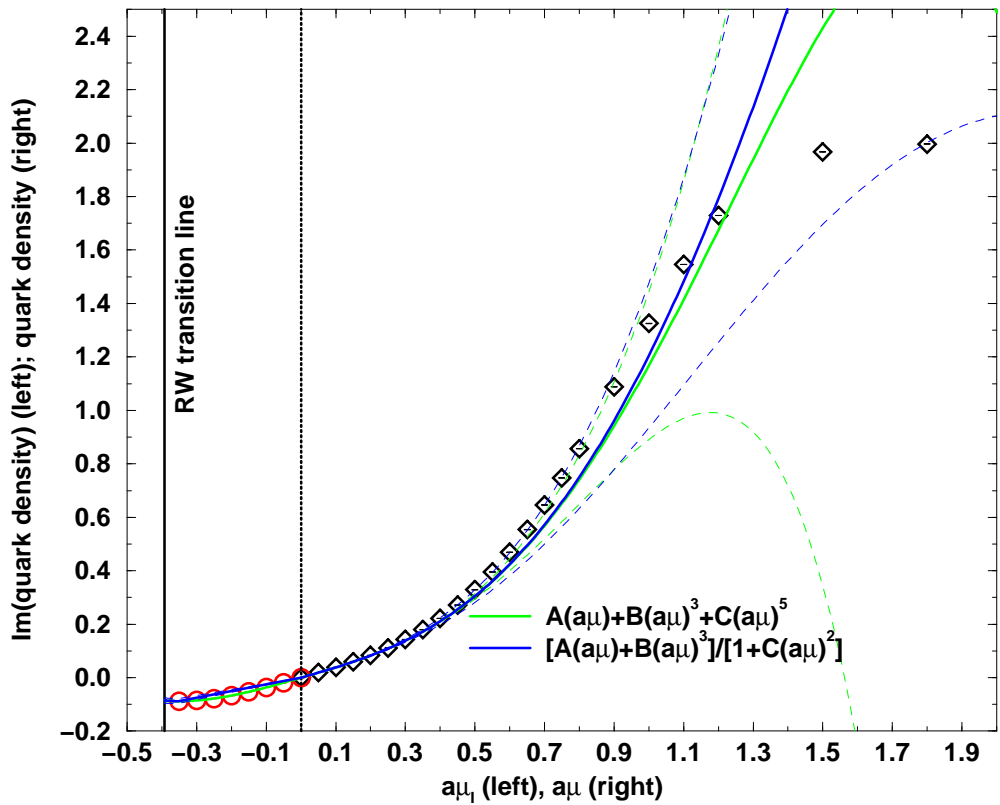}
\caption[]{Negative side of the horizontal axis: imaginary part of the fermionic number density
{\it vs.} the imaginary chemical potential. Positive side of the horizontal axis: real part of the 
fermionic number density {\it vs.} the real chemical potential. The green (blue) solid lines
represent the polynomial (ratio of polynomials) interpolating function; the dashed lines 
give the corresponding uncertainty, coming from the errors in the parameters of the fit.}
\label{fig3}
\end{center}
\end{figure}

\begin{figure}
\begin{center}
\includegraphics[width=0.75\textwidth,angle=-90]{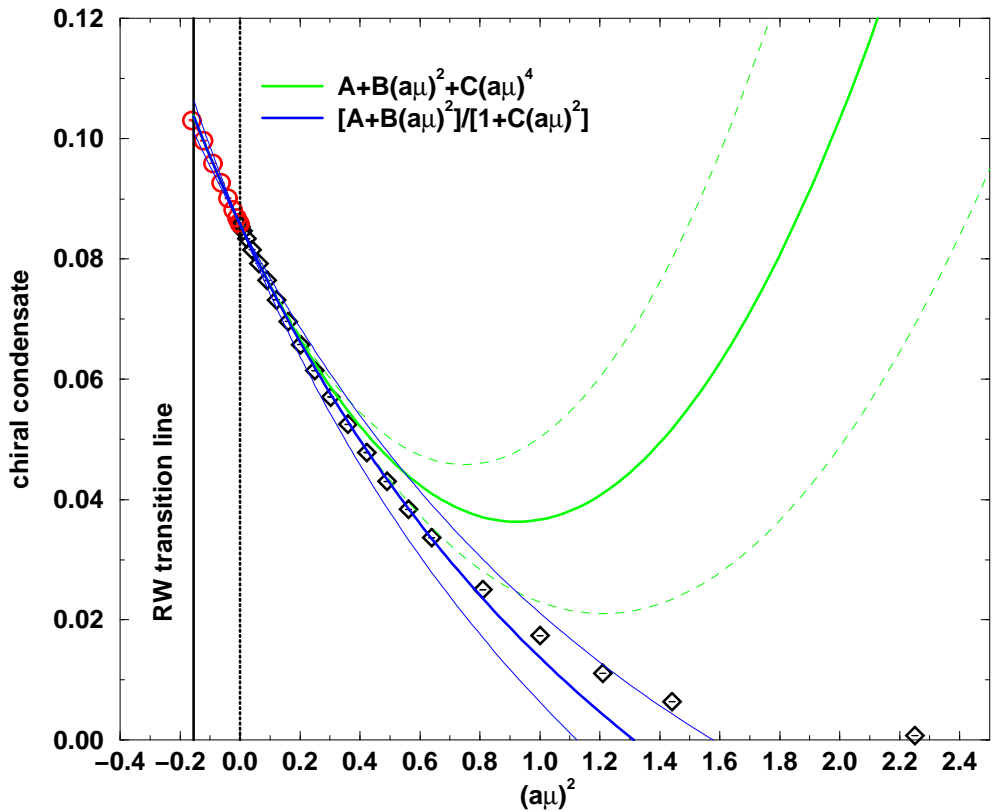}
\caption[]{Chiral condensate {\it vs.} $\mu^2$. The solid and dashed lines have the same meaning
as in Fig.~3.}
\label{fig4}
\end{center}
\end{figure}

\begin{figure}
\begin{center}
\includegraphics[width=0.75\textwidth,angle=-90]{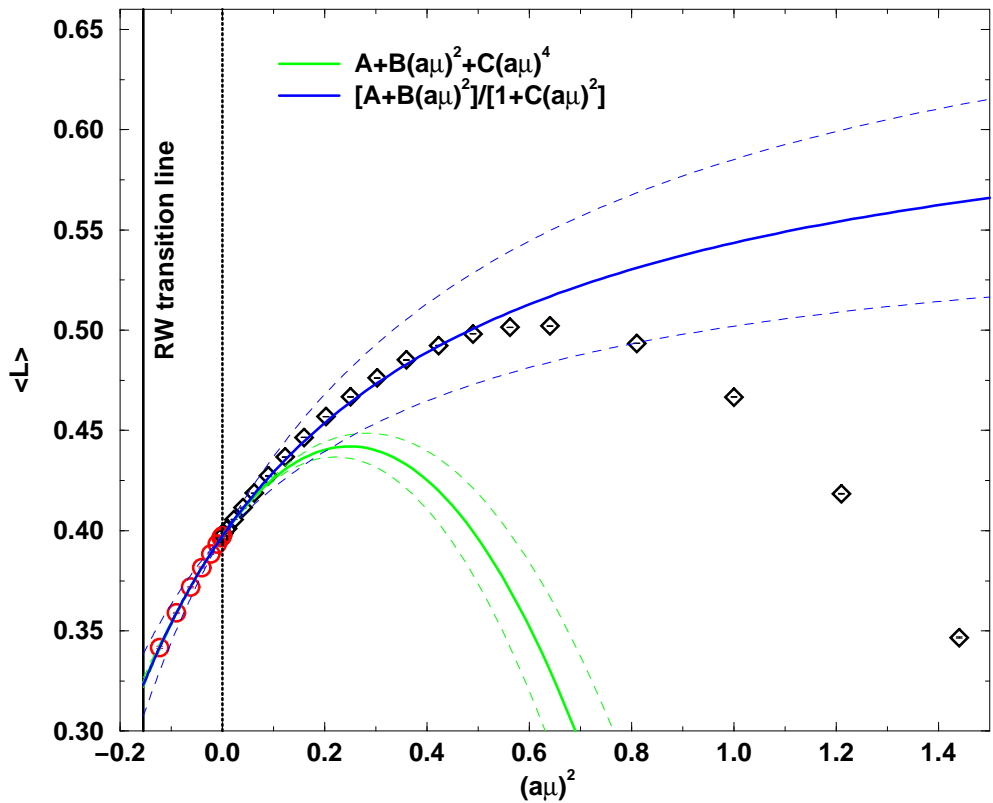}
\caption[]{Polyakov loop {\it vs.} $\mu^2$. The solid and dashed lines have the same meaning
as in Fig.~3.}
\label{fig5}
\end{center}
\end{figure}

Our findings are summarized in Figs.~\ref{fig3}, \ref{fig4} and \ref{fig5}. In Fig.~\ref{fig3}
we put on the same plot the imaginary part of the fermionic number density as a function of  $\mu_I$
and the real part of the fermionic number density as a function of the real $\mu$. The two data sets
match smoothly in $\mu=0$, which is a necessary condition for the applicability of the method of 
analytical continuation. The fermionic density approaches 2 for large values of the 
real chemical potential. This saturation 
effect is artificial and is due to the fact that no more than two fermions per site can be 
accommodated in the lattice (``Pauli blocking''). The solid lines represent the two kinds of 
interpolating 
functions, whose parameters are determined by a fit on the data at imaginary chemical potential. 
Here both interpolations, polynomial and rational function, nicely reproduce the data
at real $\mu$ over a large interval. Deviations start at values of $\mu$ for which  
the saturation effects are certainly important.

In Fig.~\ref{fig4} we show the chiral condensate as a function of $\mu^2$. Again data at imaginary 
$\mu$, i.e. $\mu^2<0$, and data at real $\mu$, i.e. $\mu^2>0$, nicely match at $\mu=0$. This time 
the different behavior of the two kinds of interpolation clearly emerges. The ratio of first order
polynomials in $\mu^2$ reproduces the data at real $\mu$ on a much larger interval than the second
order polynomial in $\mu^2$. Deviations arise for values of real $\mu$ for which saturation effects
are probably already important.
The same conclusions can be drawn from Fig.~\ref{fig5} which shows data and interpolations
for the Polyakov loop.  

The above conclusions do not change if larger order terms are included in the polynomial
interpolation~(\ref{poly}). In fact, larger order polynomials fail to reproduce the data at 
real $\mu$ even earlier in $\mu$ than second order polynomials. This is due to the fact that the
higher order terms of the polynomial are the less accurately determined in the fit 
to data at imaginary $\mu$.  On the other side, if in the ratio of polynomials
the order of the polynomials at the numerator and/or at the denominator is
increased, no improvement is observed. 

An approach alternative to that based on ratio
of polynomials is to use large order polynomials and then to build out of them rational
functions by means of Pad\'e approximants\footnote{The use of Pad\'e approximants for the 
analytical continuation of the critical line from imaginary to real chemical potential
has been suggested in Ref.~\cite{Lom05}.}. This alternative approach behaves in a way
comparable to ratio of polynomials.  

\section{Conclusions and outlook}

We have verified by comparison with direct Monte Carlo determinations
at real chemical potential in 2-color QCD that the method of analytical
continuation considerably improves if ratio of polynomials is used as interpolating 
function instead of truncated Taylor series.

In the case of the Polyakov loop and of the chiral condensate an
interpolation of numerical data at imaginary chemical potential over
the window permitted by Roberge-Weiss singularities allows an
extrapolation to real values of the chemical potential over a much larger 
region.

Deviations at very large values of the chemical potential could be due
to unphysical saturation of the fermionic density (``Pauli blocking'').
%

\end{document}